\documentclass[%
 reprint,
 amsmath,amssymb,
 aps,
]{revtex4-1}

\usepackage{graphicx}
\usepackage{bm}
\usepackage{tcolorbox} 
\usepackage{float} 
\usepackage{longtable}
\usepackage{tabularx}
\usepackage{color}
\usepackage[toc,page]{appendix}

\begin{document}

\title{Machine learning methods for locating re-entrant drivers from electrograms in a model of atrial fibrillation}

\author{Max Falkenberg McGillivray$^{1,2,\dagger, *}$}
\author{William Cheng$^{1,2,\dagger}$}
\author{Nicholas S. Peters$^{3}$}
\author{Kim Christensen$^{1,2,3}$}
\affiliation{$^1$ The Blackett Laboratory, Imperial College London, United Kingdom}
\affiliation{$^2$ Centre for Complexity Science, Imperial College London, United Kingdom}
\affiliation{$^3$ ElectroCardioMaths Programme, Imperial Centre for Cardiac Engineering, Imperial College London, United
Kingdom}
\affiliation{$^\dagger$\textbf{These authors contributed equally to this work.}}
\affiliation{$^*$\textbf{Corresponding Author: mff113@ic.ac.uk}}

\date{\today}

\begin{abstract}
Mapping resolution has recently been identified as a key limitation in successfully locating the drivers of atrial fibrillation. Using a simple cellular automata model of atrial fibrillation, we demonstrate a method by which re-entrant drivers can be located quickly and accurately using a collection of indirect electrogram measurements. The method proposed employs simple, out of the box machine learning algorithms to correlate characteristic electrogram gradients with the displacement of an electrogram recording from a re-entrant driver. Such a method is less sensitive to local fluctuations in electrical activity. As a result, the method successfully locates 95.4\% of drivers in tissues containing a single driver, and 94.8\% (92.5\%) for the first (second) driver in tissues containing two drivers of atrial fibrillation. Additionally, we demonstrate how the technique can be applied to tissues with an arbitrary number of drivers. Extending the technique for use in clinical practice could alleviate the limitations in current ablation techniques that arise from limited mapping resolution. 

\vspace{3.5mm}

\noindent \textbf{Keywords:} atrial fibrillation, arrythmia, cellular automata, targetted ablation, machine learning, electrograms
\end{abstract}

\maketitle

\section{\label{sec:level1A}Introduction}

Atrial Fibrillation (AF) is the most common cardiac arrhythmia in clinical practice and is getting more prevalent in the general population due to the aging demographic. However, the mechanistic origin of AF is still poorly understood. As a result, the success rates of treatment options remain limited with future improvements requiring a better understanding of how AF emerges from the underlying properties of the myocardium.

A variety of possible mechanisms have been proposed to explain the origin of AF. These include circus movement re-entry, the leading circle theory, spiral wave re-entry (otherwise known as rotors) of which micro-anatomical re-entry can be thought of as a subset, and the multiple wavelet hypothesis \cite{nattel2002,schotten2011}. However, there are many contradictory findings, and no one mechanism explains all observations \cite{nattel2017,lee2015}.
This suggests new techniques are needed both in clinical practice and research, with numerous researchers highlighting the potential of computational simulations and machine learning \cite{Trayanova2017,clayton2011,hood2004,winslow2012,deo2015,li2014}. 
An issue of particular importance is that of limited mapping resolution when detecting the drivers of AF. Errors in the accuracy of imaging data limit the efficacy of treatment by ablation, and it cause disagreement when interpreted by the research community \cite{Roney2017,hansen2015,cherry2008}.

In this paper, we present a method whereby electrograms are extracted from a number of independent locations in the atria and these are cross-referenced to triangulate the position of a re-entrant circuit. By using multiple measurements, noise and local fluctuations are minimised and a prediction can be reached without being overly reliant on the imaging resolution at any one given location. The procedure applies machine learning methods to maximise the prediction accuracy of the algorithm. The work here should be considered a proof of concept and has been carried out without some of the detail that would be necessary in a realistic clinical implementation, but nevertheless, it presents a clear path towards a potential improvement in ablation success rates based on locating drivers from electrograms only.

The Christensen, Manani and Peters model (CMP) is a 2D cellular automata on a simplified architecture of the atria \cite{christensen2015}. While the model architecture is simplified, it preserves the key features of discrete cardiomyocyte activation at the microscopic level, while ensuring macroscopic conduction appears continuous. The CMP model has a particular focus on demonstrating the role of fibrosis in initiating and maintaining AF. AF is driven through the spontaneous emergence of re-entrant circuits which generate rapid, irregular atrial activity. This mechanism is a form of micro-anatomical re-entry. These circuits form at regions with high levels of localised fibrosis. The model also explains the wide range of AF classifications from short intermittent episodes to long lasting permanent AF as increasing levels of fibrosis modify the myocardium \cite{manani2016}. In addition to the mechanistic benefits of cellular automata, their popularity has recently grown given their capacity to simulate much longer timescales of atrial activity than computationally intensive continuous models \cite{lin2017,lord2013,ciaccio2017b}.  

Machine learning approaches are statistical techniques implemented computationally which excel at finding hidden insights in complex data without being explicitly programmed to do so.  In particular, machine learning has had recent successes in the medical community in classifying skin melanoma and in genetic sequencing \cite{Esteva2017,Libbrecht2015}. These methods often rely on having a large dataset for the models to learn correlations from. In clinical medicine, the lack of good quality data in large quantities often makes such an approach difficult. However, when working with computer simulations such as those in the CMP model, data can be generated in sufficient quantities, and hence, the statistical insights these models provide can offer significant improvements when analysing data. The computational complexity of continuous models make these unsuited to generating large quantities of data \cite{lin2017,clayton2011}.

The remainder of this paper is organised as follows. First, we briefly introduce the CMP cellular automata model used in this research. We describe the physiological motivation behind the model and outline the process of simulating electrograms. We discuss a novel visualisation of electrograms and use insights from this method to inform statistical analysis. Random Forests, a machine learning technique with consistently strong results across a number of domains, are described briefly in section \ref{sec:level1B} \cite{Caruana2006,caruana2008}. These are then implemented in a search algorithm for locating AF drivers. Our results are presented and discussed in section \ref{sec:level1C}. 
Finally, we outline potential future work and discuss the CMP model's relevance to current clinical research.

\section{\label{sec:level2B} The Christensen--Manani--Peters model}

The Christensen--Manani--Peters (CMP) model is a 2D cellular automata. Each atrial muscle cell in the CMP model is represented by a single square in a larger square lattice of side length $L$. All cells are connected to their nearest neighbours in one dimension (the longitudinal direction) and in the orthogonal dimension with probability $\nu$ (stochastic connections, the transverse direction). This construction is a simplified computational implementation of the real myocardium, but it preserves the essential myocardial architecture which ensure that electrical impulses travel preferentially along muscle fibres rather than laterally between fibres. The strong coupling in the longitudinal direction represents individual muscle cells forming a single, long muscle fibre. The reduced coupling in the transverse direction represents the reduced electrical conductivity between neighbouring muscle fibres. The parameter $\nu$ controls the strength of the transverse coupling. This is manifested in the anisotropy of electrical conduction velocities. There are periodic boundary conditions in the transverse direction (across different muscle fibres) and open boundaries in the longitudinal direction (along a single muscle fibre) representing a simplified cylindrical geometry of the atria \cite{christensen2015}. The cells along the left boundary are the pacemaker cells which are excited every $T$ time steps. 

In the CMP model, the electrical cycle of a given cell is as follows: Cells are updated in discrete time steps. Initially a cell is at rest. In this state, the resting cell can become excited at the next time step by a neighbouring excited cell. The cell is in the excited state for one time step after which the cell is in the refractory state for the next $\tau$ time steps. In the refractory state, a cell is unexcitable after which the cell re-enters the resting state. This cycle is indicative of the action potential of real cardiomyocytes. The beating of the atria is simulated by exciting all the cells along the left boundary of the tissue. This signal can then propagate across the tissue until it is dissipated at the opposite open boundary. The pacemaker cells are activated every $T$ time steps. The full excitation rules of the CMP model are summarised in the box below.

\begin{tcolorbox}[width=\linewidth, title = \textbf{CMP Algorithm}]
\begin{enumerate}\itemsep5pt
\item All cells are connected longitudinally forming a muscle fibre.
\item A fraction $\nu$ of the transverse connections between cells are filled linking two muscle fibres. 
\item A resting (excitable) cell connected to an excited cell at time step $t$ will become excited at time step $t+1$.
\item A refractory (unexcited) cell connected to an excited cell at time step $t$ will not become excited at time step $t+1$.
\item A cell remains refractory for a duration of $\tau$ time steps before returning to the resting state.
\item Every $T$ time steps, the pacemaker cells along one of the open boundaries of the muscle tissue are excited.
\item A small fraction $\delta$ of cells are dysfunctional and have a small probability $\epsilon$ of not getting excited.
\end{enumerate}
\end{tcolorbox}

\begin{figure}
    \centering
    \includegraphics[width = \linewidth]{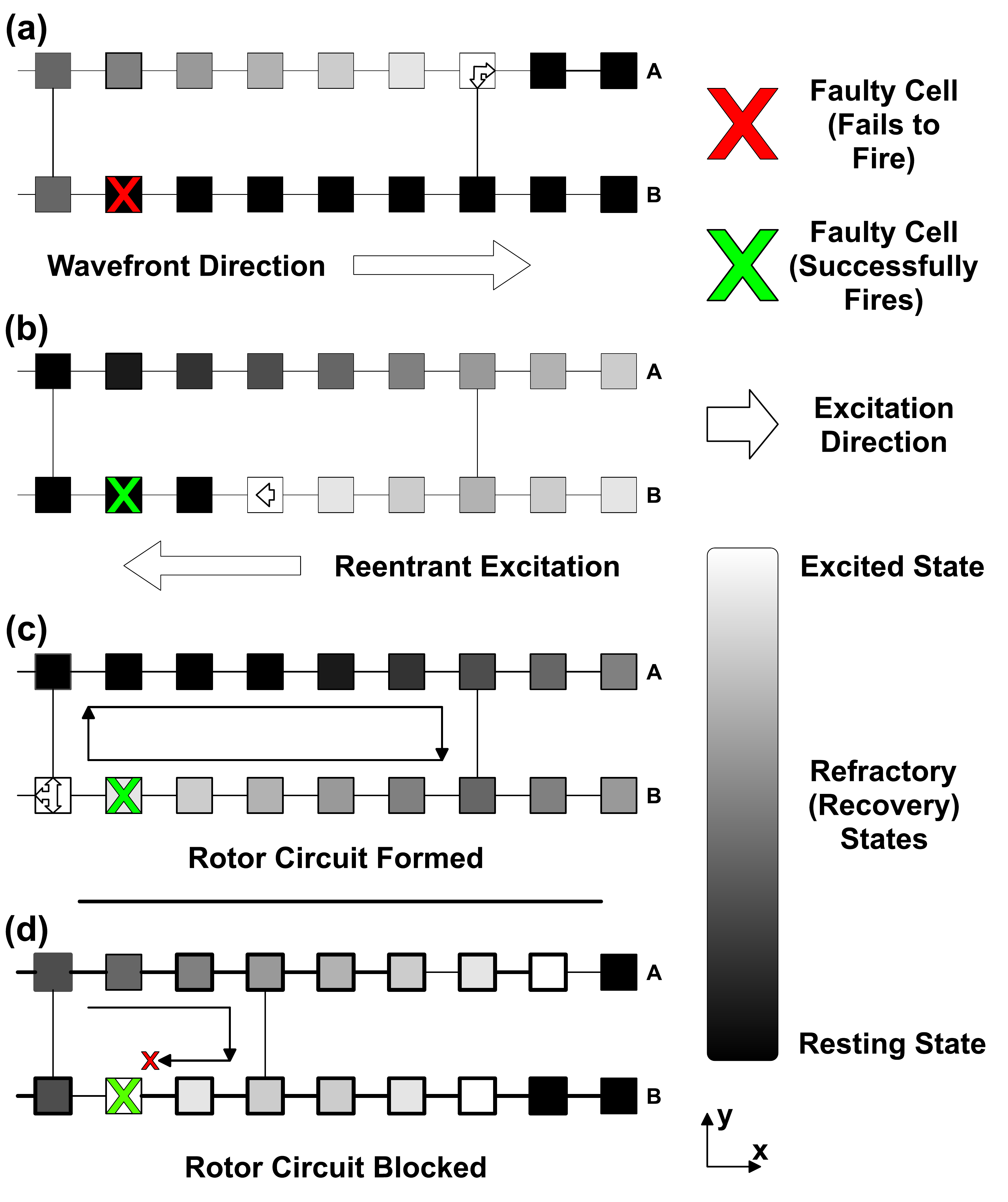}
    \caption{The formation of a simple re-entrant circuit at the cellular level. Each cell corresponds to a single muscle cell. An excited cell is shown in white, resting (excitable) cells are black. Cells shown in grey are refractory (unexcitable) for the duration of the refractory period. All cells are coupled to their nearest neighbours longitudinally -- this reflects the strong coupling of cells along muscle fibres. Transverse couplings exist with probability $\nu$ -- reducing $\nu$ reflects the decoupling between adjacent muscle fibres caused by fibrosis. An excitation is initiated along the left wall of the heart tissue and propagates left to right. When the excitation reaches a dysfunctional cell which fails to fire (marked by a red cross), \textbf{(a)} the excitation in fibre \textbf{B} is blocked but excitations continue in fibre \textbf{A} above. When a coupling between the excited and blocked strands is reached the excited cell can send a signal propagating backwards from right to left down the blocked strand \textbf{(b)}. If the path length of the re-excitation is sufficiently long, the re-entrant excitation can excite tissue behind the main wavefront. This signal can then move to the adjacent strands forming a continuosly re-excited circuit. The simple re-entrant circuit shown here is rectangular in shape and is formed from two fibres (\textbf{A} \& \textbf{B}) and a single dysfunctional cell. More complicated re-entrant circuits can consist of multiple fibres and multiple dysfunctional cells. The extract shown in \textbf{(d)} is for a different region of heart tissue. Here the same mechanism as above attempts to form a re-entrant circuit. However, the circuit path length is insufficiently long such that the re-entrant excitation is blocked by the refractory cell to the left of the dysfunctional cell marked with a cross. Hence, a continuously excited circuit cannot form in this tissue segment. This behaviour explicitly links the emergence of re-entrant circuits with regions of high fibrosis (low $\nu$).}
    \label{fig:rotorform}
\end{figure}

\begin{figure}
\centering
\includegraphics[width=\linewidth]{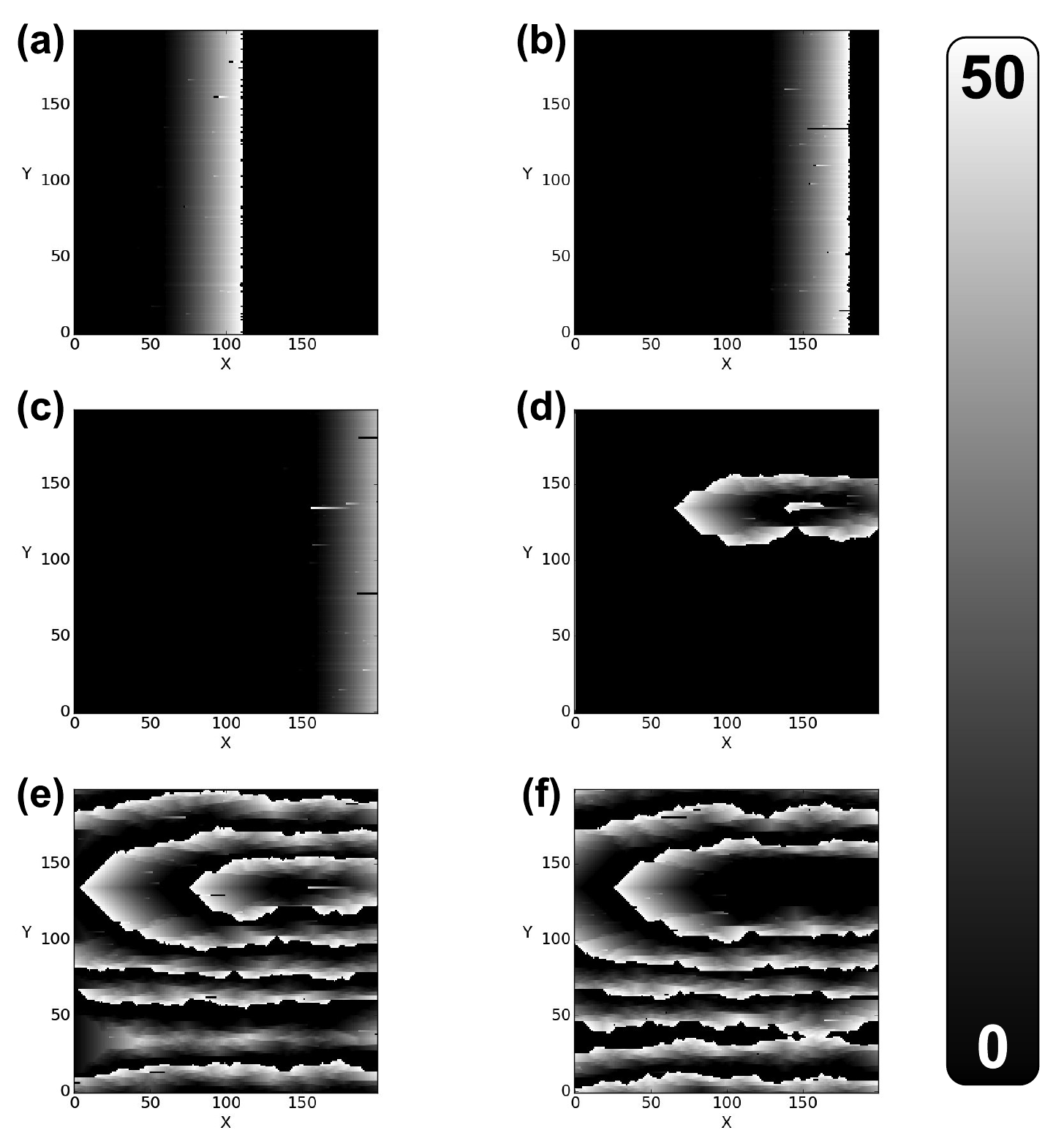}
\caption{The formation of fibrillatory re-entrant circuits in the Christensen-Manani-Peters model for a refractory period of $\tau = 50$. Excited cells are shown in white, resting (excitable) cells are black and cells which are refractory (unexcitable) are shown in grey. \textbf{(a)} A planar wave initiated by the pacemaker cells at the left boundary of the cylindrical heart tissue propagates from left to right along muscle fibres. \textbf{(b)} A dysfunctional cell fails to fire blocking the propagation of the signal along that fibre. Excited cells in the adjacent muscle fibre can re-excite the blocked fibre at the next vertical coupling, see Fig. \ref{fig:rotorform}. \textbf{(c)} If the pathlength of the re-excitation is sufficiently long, the re-entrant excitation can escape the unexcitable region. \textbf{(d)} This escaped excitation initiates a continuously activated circuit from which chaotic waves propagate -- this circuit is the driver of AF. \textbf{(e)} Fibrillatory waves spread across the whole tissue and prevent the formation of new planar wavefronts at the pacemaker cells on the left boundary. \textbf{(f)} If a dysfunctional cell in the re-entrant circuit fails to fire the circuit is stopped and fibrillatory behaviour dissipates.}
\label{fig:propagation}
\end{figure}

Dysfunctional cells can block propagation along the cell strands and leave an opening for the propagating wavefront to turn back in on itself forming a circuit (re-entry). If this circuit is long enough (which occurs when $\nu$ is sufficiently small), the signal can continuously propagate around the circuit forming a persistent driver of AF, see Fig. \ref{fig:rotorform}. Figure \ref{fig:propagation} shows the emergence of AF on a tissue wide scale. In the real myocardium, the transverse decoupling of muscle fibres is associated with the build up of fibrosis \cite{zahid2016,haissaguerre2016,matsuo2009}. Hence, the parameter $\nu$ can be thought of as a control for the degree of fibrosis in the myocardium. In previous work with the CMP model, we have shown how the increased prevalence of fibrosis (decreasing $\nu$) increases the risk of AF persisting, in agreement with clinical observation \cite{christensen2015,manani2016}.

It is helpful to understand the considerations behind the dysfunctional cell mechanism. The important detail is that the AF mechanism in the CMP model is possible using any mechanism of unidirectional conduction block - the stochastic failure of cells to depolarise is a simple way to include this effect in the model. In the original CMP model, the $1000 \times 1000 $ grid is coarse grained into a $200 \times 200$ grid for computational ease. A $1000 \times 1000 $ grid would approximately account for the total number of cardiomyocytes on the epicardial surface of the atrium. The model then treats each cell in the coarse grained grid as a single individual muscle cell. For single cardiomyocytes in isolation there is little clinical evidence for the cell stochastically failing to depolarise. However, instead of considering the CMP model as a microscopic depiction of atrial conduction we can think about a mesoscopic picture where each cell in the $200 \times 200$ grid represents the average dynamics of the underlying $5 \times 5$ block of cells. These coarse grained blocks can still follow the same branching/connectivity rules between cells as originally formulated for the CMP model. 

Within this $5 \times 5$ cell block, we can consider the possibility of unidirectional conduction block due to the imbalance between current sources and sinks - the possibility of such a mechanism has been shown by \citeauthor{bub2002} \cite{bub2002} in a theoretical model and has also been supported by clinical evidence \cite{fast1995,ciaccio2016, ciaccio2017}. This explains why we might expect any given block of cells to display unidirectional conduction block with some small probability, $\epsilon$, given a suitable geometric arrangement of cells with a small probability, $\delta$. It is also sensible to consider such an effect to be stochastic since leaking current over a number of activation cycles can push a previously blocked group of cells over the depolarisation threshold.

The parameters used in this work are $\nu = 0.2$, $\tau = 50$ and $T = 220$. We fix the coupling fraction to be $\nu = 0.2$ since this is the largest coupling fraction at which we typically observe paroxysmal AF \cite{christensen2015,manani2016}. The shortest timescale in the CMP model is associated with the excitation timescale of a single cardiomyocyte of $\Delta t = 0.6 \text{ms}$. In the course grained tissue, this corresponds to an excitation timescale for each cell of $\Delta t^{*} = 5 \cdot 0.6 \text{ms} = 3.0 \text{ms}$. The time step of $3ms$ approximately corresponds to the time taken for an electrical signal to cross a $5 \times 5$ block of cardiomyocytes in the real atrium along a muscle fibre. The refractory period is chosen to be $\tau = 50$ time steps where the unit of time is $\Delta t^{*}$ such that the action potential of a cardiomyocyte is $150 \text{ms}$ in accordance with the values seen for human atrial cardiomyocytes. In sinus rhythm the CMP model therefore beats approximately once every $660 \text{ms}$. Note that AF can easily be induced with a different set of parameters, but these are chosen as a realistic reflection of atrial activity. In the original CMP model, the fraction of dysfunctional cells is $\delta = 0.05$. This is an arbitrary choice which allows AF to spontaneously emerge. In this work we set $\delta = 0$ and instead artificially insert re-entrant circuits into the tissue. This is a reasonable concession for computational ease since this work is not concerned with the spontaneous emergence of AF but rather the diagnosis of AF after it has emerged.   

\section{\label{sec:level1B}Methods}

To locate AF drivers on the CMP tissue, machine learning models are utilised. These models are trained with electrogram feature data gathered from a large number of simulations of varying AF instances on CMP tissue. A benefit of this approach is that instead of relying on a single accurate measurement to locate a re-entrant circuit, a collection can be used to determine the circuit's location. This means that the electrograms can be measured at a lower resolution as no one measurement is critical to the success of the locating algorithm. Sample electrograms from the CMP model can be seen in Fig. \ref{fig:comparisonECG}. These are generated as outlined in appendix \ref{sec:ap1}.

\begin{figure}[hb!]
\centering
\includegraphics[scale = 0.51]{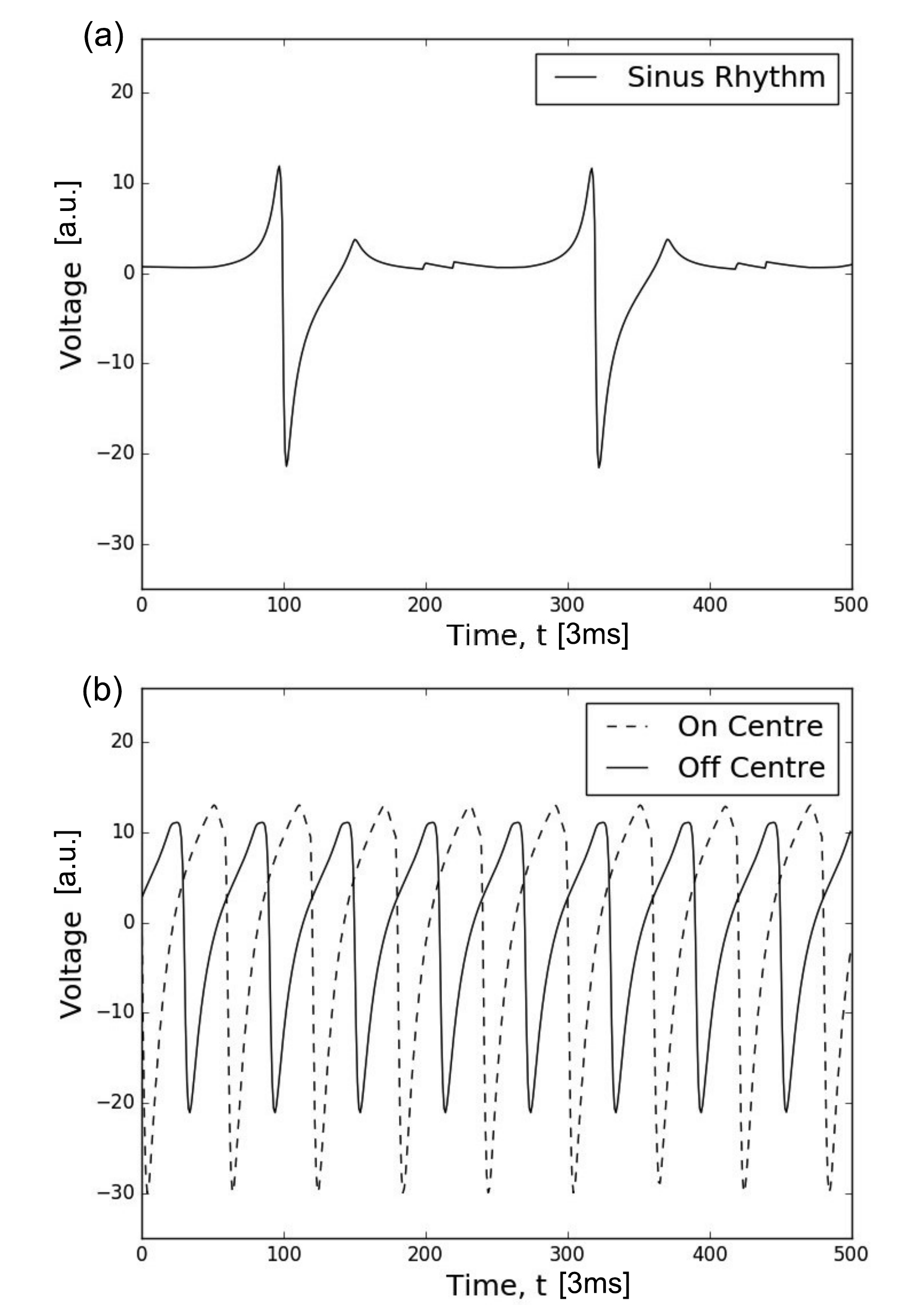}
\caption{Electrograms simulated on perfect isotropic tissue ($\nu = 1.0$) in the CMP model using Eq. \eqref{eqn: ECG}. Voltage is measured in arbitrary units. Time is measured in steps of 3ms such that at sinus rhythm the pacemaker cells activate every 660ms (220 time steps). \textbf{(a)} An electrogram recorded during regular sinus rhythm. The large depolarisations at approximately $t = 100$ and $t = 320$ correspond to the planar wavefront crossing the electrogram recording probe. The small fluctuations at approximately $t = 200$ and $t=420$ correspond to the activation of a new planar wavefront at the left boundary by the pacemaker cells and the dissipation of electrical activity at the open boundary on the right after the wave propagates through the tissue -- in this sense these are small finite size effects. \textbf{(b)} Electrograms recorded during the rapid pacing of the heart where electrical wavefronts originate from a single point in the tissue. The dashed and solids lines show electrograms recorded at the centre of the electrical activity and thirty cells displaced from the centre, respectively. There are clear visual differences between electrograms recorded at different locations relative to the centre of electrical activity. Note, the clear visual difference between electrograms is significantly less pronounced in imperfect, anisotropic tissue ($\nu < 1.0$) -- this warrants the use of statistical techniques for analysis.}
\label{fig:comparisonECG}
\end{figure}

\begin{figure}
\centering
\includegraphics[width = \linewidth]{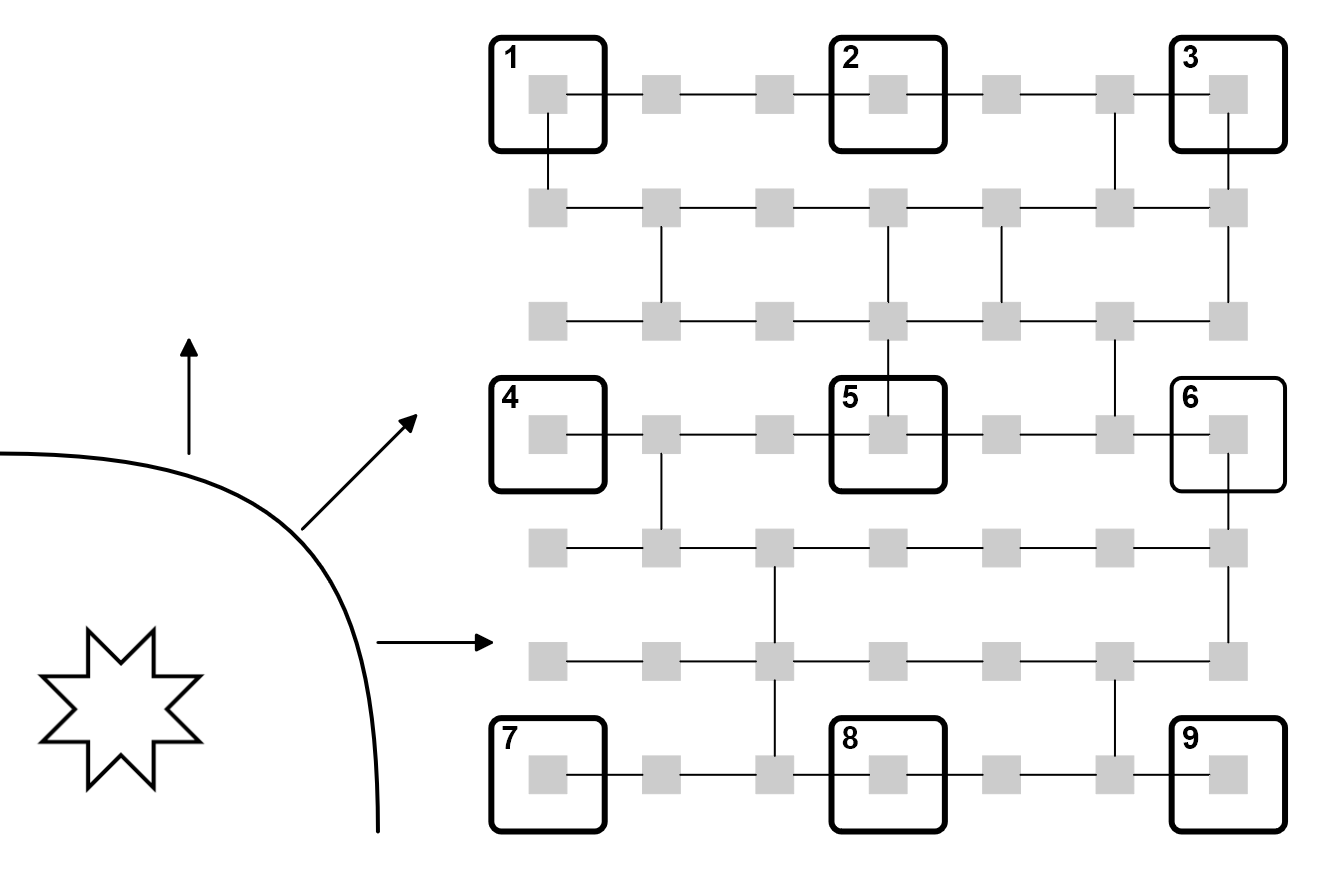}
\caption{Electrical wavefronts propagate in all directions from an AF driver (shown as the star in the bottom left). A $3 \times 3$ electrode grid is used to record 9 simultaneous electrograms (large boxes numbered 1 to 9) across a region of $9 \times 9$ cells in the CMP model (small squares shown in grey). By spacing out the electrodes, the approaching electrical wavefront will cross each of the electrodes in a slightly different order and in a slightly different direction affecting the feature behaviour seen at each electrode. Hence, taking the gradient of features across the electrode grid can give detailed information about the wavefront flow across the grid enhancing our knowledge of the driver's position.}
\label{fig:elecgrid}
\end{figure}

For the purpose of recording training data to build the machine learning models, a 9 electrode probe used, arranged in a $3 \times 3$ array. This represents the multi-contact mapping catheters used for ablation. The $3 \times 3$ probes are distributed uniformly in a $9 \times 9$ cell grid of the CMP tissue as seen in Fig. \ref{fig:elecgrid}. The distance between each probe and its nearest neighbors is 3 cells as this gives a resolution comparable to clinically used ablation mapping catheters. We define the probe to be on the re-entrant circuit if any part of the re-entrant circuit is within the multiprobe's $9 \times 9$ cell region.

A first look at the electrograms shows clear distinctions depending on the electrodes relative position from the re-entrant circuit as shown in Fig. \ref{fig:comparisonECG} (b). Features gathered from these individual electrodes include: maximum voltage, minimum voltage, first stationary point position, mean voltage, waveform skewness and other common statistical measures. A full list can be found in appendix \ref{sec:ap2}. With the 9 probe array, the wavefront from the AF circuit will reach each individual electrode at a different time. Therefore, gradients of the features can be measured across the whole electrode array in different directions giving more effective information on the position of the AF circuit, see Fig. \ref{fig:elecgrid}. 

\begin{figure}
\centering
\includegraphics[width=\linewidth]{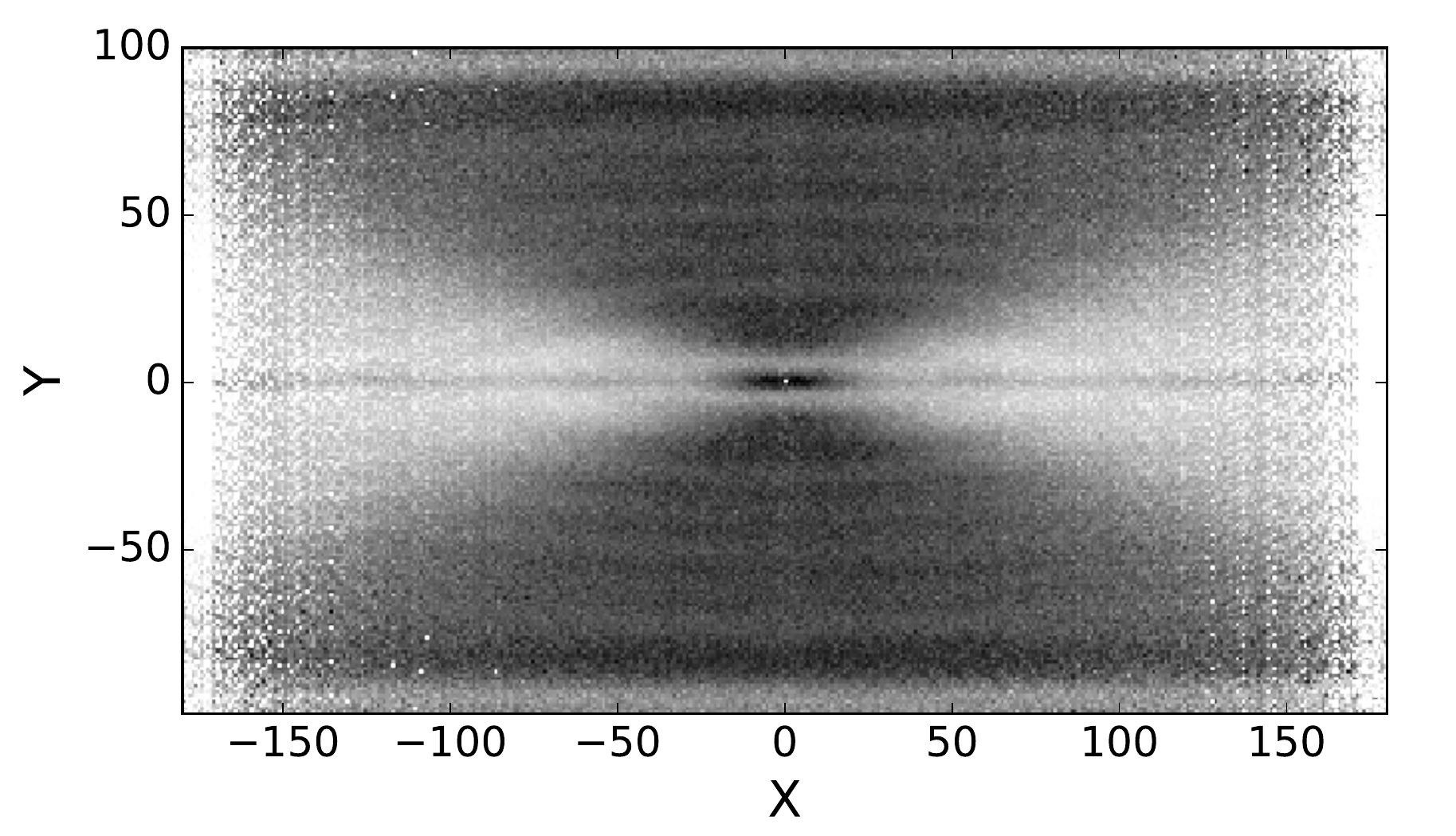}
\caption{The vector feature map for the dominant frequency of the electrograms Fourier transform on an arbitrary scale, shown by the greyscale. The image is generated from tissues with a single ectopic cell beating every 60 time steps placed at a random location in the tissue with transverse coupling fraction $\nu = 0.2$. The image shows strong separation between regions of unidirectional wavefront propagation (light grey) and bidirectional wavefront propagation (dark grey). There is also a strong indicator of the driver's centre, black region at (0,0). Notice that the image is symmetric across the X and Y axes and that different feature maps highlight different wavefront dynamics. There is significant differentiation between bulk regions but little differentiation in small, localised regions. Combining multiple vector feature maps can improve local differentiation.}
\label{fig:FCplot_STATIONARY}
\end{figure}

Visualising the features from electrogram data can give significant insight into the electrical dynamics of the CMP model without extensive statistical analysis. This can be done using a visualisation we have coined the vector feature map. Their creation, analysis and general features are described in detail in appendix \ref{sec:ap2}. A vector feature map shows the average value of a given vector relative to the centre of the re-entrant circuit in many different instances of the CMP model. An example is shown for the magnitude of the dominant Fourier transform frequency of the electrogram signal in Fig. \ref{fig:FCplot_STATIONARY}.

To locate AF drivers using electrogram features, a supervised machine learning technique known as the Random Forest model is used \cite{breiman2001}. The model is capable of giving quantitative (the distance between the probe and the re-entrant circuit) and qualitative (the probe is/is not currently on the re-entrant circuit) responses when given a set of electrogram feature information \cite{james2013introduction}. The Random Forest model was chosen due to its relative effectiveness compared to other machine learning models. The method has recently been used for problems such as  tissue segmentation in the brain and the classification of heart failure subtypes \cite{pereira2016, Austin2013}. \citeauthor{Caruana2006} \cite{Caruana2006,caruana2008} note that the Random Forests model has one of the highest average performances of any machine learning method across a wide range of different problems. Preliminary testing on our data showed considerably higher success rates for Random Forests than for other simple machine learning algorithms. The mathematical background for Random Forests is described in appendix \ref{sec:ap3}.

The training data for the Random Forests was gathered from 5000 randomly generated CMP tissues with one randomly placed AF circuit. The fraction of transverse connections was chosen to be $\nu = 0.2$ as this is the critical point where instances of paroxysmal AF emerge in the CMP model \cite{christensen2015}.  Each tissue had 64 multiprobe electrodes uniformly placed giving in total of 2,880,000 electrogram recordings in total.

\subsection{\label{sec:level2E} Algorithm for locating re-entrant circuits}

The goal of our algorithm is to demonstrate a proof of concept where re-entrant circuits driving AF can be located using solely electrogram information. The aim is not to create a perfect model which could be directly transfered to more complicated scenarios, but rather, the aim is to show the feasibility of these methods for electrical mapping in a system with large local fluctuations. As part of this approach, a small number of simplifications are applied to the CMP model to simplify simulations and analysis.

All tissue instances are generated at a fixed transverse coupling fraction of $\nu = 0.2$ which is approximately the degree of fibrosis at which we observe paroxysmal AF. We also work in the low noise limit where $\delta = 0$. As a result, temporary critical circuits cannot spontaneously form in the tissue as shown in part (d) of Fig. \ref{fig:rotorform}. The limitations of this decision are discussed in section \ref{sec:lims}. The low noise limit inhibits the formation of new, stable re-entrant circuits. Instead, a single circuit is artificially constructed by picking a random point in the CMP tissue and removing the coupling to the muscle fibre above and below of the 28 cells to the right of the randomly chosen cell. This gives a rectangular circuit of path length $60$ -- this is an arbitrary choice which is slightly larger than the refractory period of $\tau = 50$. The circuit is then artificially forced to start driving AF by the introduction of a single ectopic beat in the circuit.

\begin{figure}
    \centering
    \includegraphics[width = \linewidth]{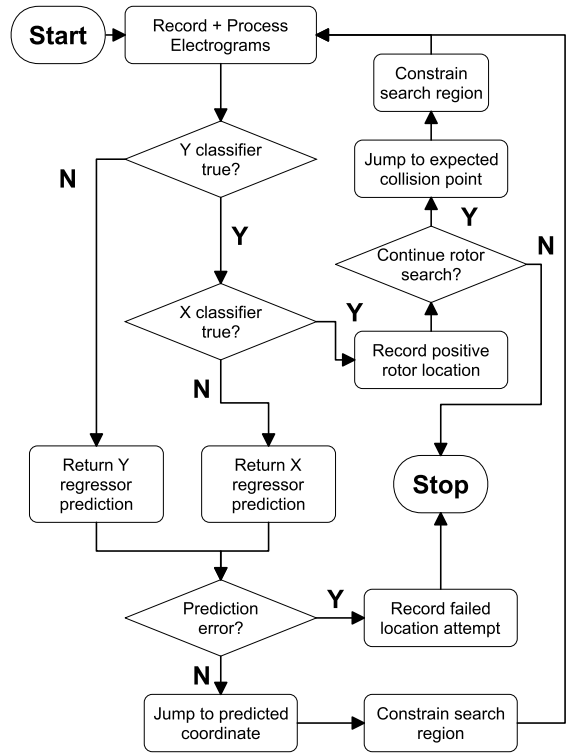}
    \caption{A flow chart describing the basic AF driver location algorithm. When starting the algorithm, the simplified CMP model is run with one or two drivers placed at random locations -- the algorithm could easily be extended to more drivers if desired as outlined in Fig. \ref{fig:multisearch}. A random position is chosen to record the first set of electrograms. Search regions are constrained as described in appendix \ref{sec:ap4}. Predictions from the $X$ and $Y$ regressors are forced to lie within search constraints. ``Prediction error'' initiates if the constraint region becomes too small or predictions loop between repeated coordinates.}
    \label{fig:alg}
\end{figure}

The procedure to be tested is split into two sections. First, the CMP tissue must be initialised to generate fibrillatory behaviour. Once initialised, the data aquisition and processing cycle can be initiated. The initialisation phase is outlined as follows and corresponds to the `Start' block in Fig. \ref{fig:alg}.

\begin{enumerate}
\item Generate an instance of the CMP model at $\nu = 0.2$ of linear size $L = 200$. Do not create any dysfunctional cells which could generate unexpected re-entrant circuits (i.e. $\delta = 0$).
\item At a random location in the tissue, insert a single simple re-entrant circuit consisting of two cell strands.
\item Allow fibrillation to commence and continue until the whole tissue has been reached by the electrical activity from the re-entrant circuit.

\end{enumerate}

After the CMP model is initialised, the data acquisition and processing cycle can be used to generate predictions of the expected location of re-entrant circuits in the tissue. This is illustrated in Fig. \ref{fig:alg} and broadly follows the following steps below. As can be seen from the lack of local differentiation in the electrogram feature in Fig. \ref{fig:FCplot_STATIONARY}, it is a non-trivial problem whether or not electrogram dynamics can be used to infer the position of re-entrant circuits drivers which motivates the use of a recursive method involving multiple measurements.

\begin{enumerate}
\setcounter{enumi}{3}
\item At a random location in the tissue, place a $3 \times 3$ array of electrode probes and generate electrograms at each position.
\item Extract statistical information from the electrograms and pre-process data for compatibility with required machine learning data structures.
\item Process data using machine learning models to output a prediction for the expected electrogram location. \label{point1}
\item Calculate the constraints of possible prediction locations based on the mechanism shown in appendix \ref{sec:ap4}.
\item Post-process prediction data to abide by the calculated constraints for the final prediction using the mechanism shown in appendix \ref{sec:ap4}.
\item Record a new set of electrograms at the predicted position of the re-entrant circuit. If the electrogram behaviour is consistent with the expected statistical features at re-entrant circuit, either end the algorithm or proceed to searching for any remaining circuits. If the position is not consistent with the expected behaviour of a re-entrant circuit, return to step \ref{point1} and repeat the prediction process.
\end{enumerate}

The algorithm utilises four Random Forest models, two classification and two probabilistic. The classification models are used to check if the probe is positioned on the drivers $X$ and $Y$ axes. The responses for these two probabilistic models are the probabilities of the driver lying on each transverse cell column on the $X$ axis and each longitudinal cell strand on the $Y$ axis -- this is an adaptation of the regression style models typical in Random Forests in which a continuous scale is broken into discrete ranges and each of the ranges is considered its own class. The benefits of the probabilistic approach over typical regression models is that it is easier to implement additional levels of post-processing in any predictions.
The probability of the re-entrant circuit lying in each class can then be processed to infer the predicted position of the re-entrant circuit in a given direction. The probabilistic models can be made from classification (qualitative) models where instead of the majority rule, the response is given by
\begin{equation}
P_i = \frac{k_i}{k}, \text{ with} ~ k = \sum_{i=1}^K k_i,
\end{equation}

\noindent where $P_i$ is the probability for the particular response $i$, $k_i$ is the number of data samples that have response $i$ and $k$ is the total number of responses. These probabilities do not consider that electrogram measurements may have been taken previously which gave information as to the direction of the re-entrant circuit from particular regions in the tissue. The post-processing procedures described in appendix \ref{sec:ap4} account for this.

\begin{figure}
    \centering
    \includegraphics[width= \linewidth]{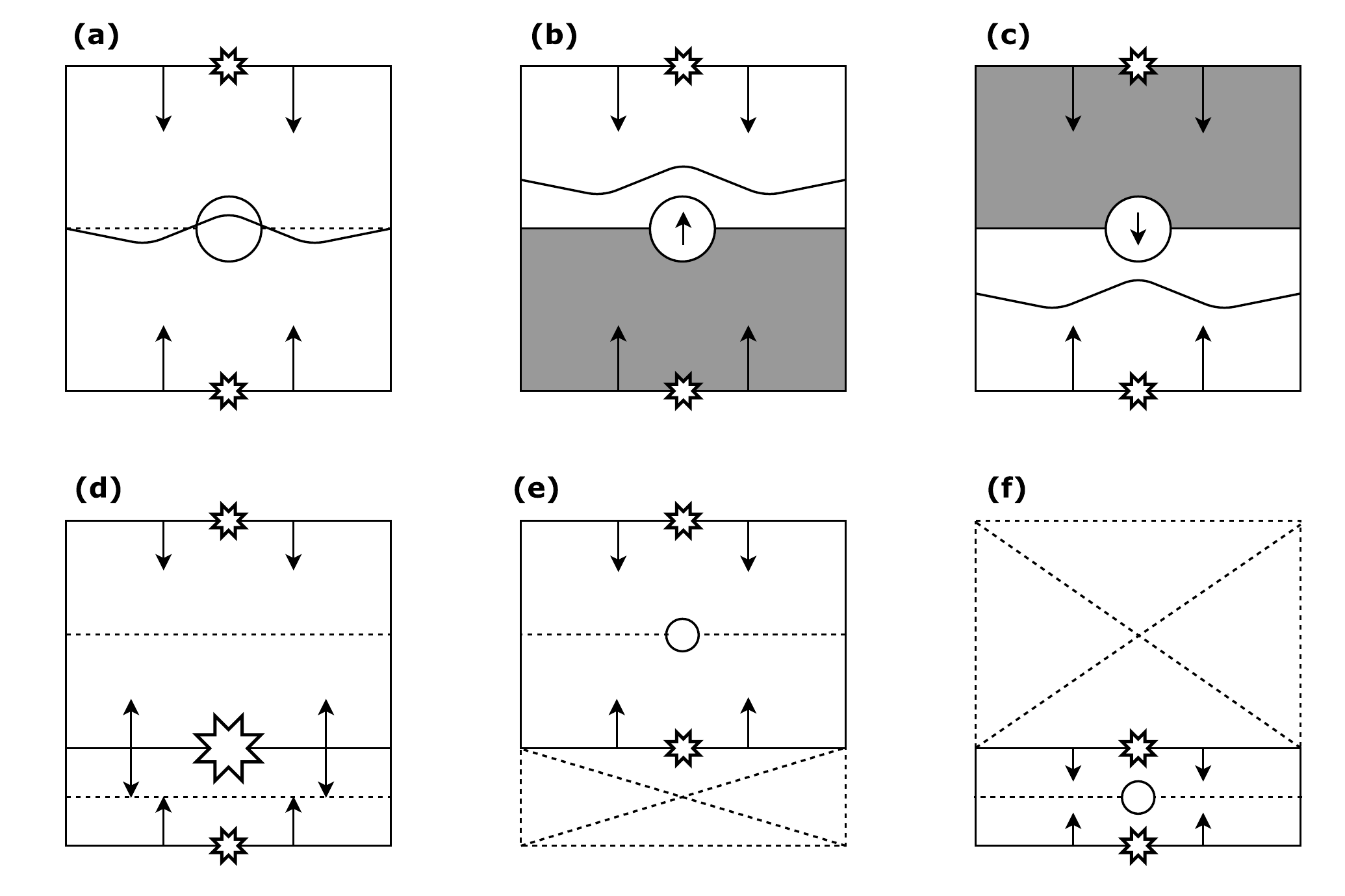}
    \caption{A diagram highlighting the key mechanism behind the extension of a single driver location algorithm to an arbitrary number of drivers. \textbf{(a)} A single driver tissue where the driver has been placed at the circular boundary to indicate wavefronts propagating from above and below (small stars). \textbf{(b/c)} The grey regions indicate the search region for a second driver inferred by observing the absence of wavefront collisions at the expected position. \textbf{(d)} A second driver is found (big star). This two driver system can be conceptually split into two single driver tissues, \textbf{(e/f)}. These can be searched using the same recursive strategy starting at \textbf{(a)}. Note that in the current work, the mechanism for finding multiple drivers is simplified to only predict the position of expected collisions in one dimension, however, this could easily be extended to the full geometry in future work.}
    \label{fig:multisearch}
\end{figure}

In a general case, it may occur that there are multiple re-entrant circuits present in a single tissue. To test our methods for this scenario we repeated the procedure outlined at the start of this section with a simple change that two re-entrant circuits are randomly placed in the tissue (with a minimum separation of 10 cells vertically to avoid overlap). We then adjust the search algorithm to start looking for a second re-entrant circuit after the first is found. However, note that the machine learning models used are still only trained on simulations with a single re-entrant circuit -- the flexibility of our model to search for multiple drivers without explicitly learning to do so is one of the major benefits of our approach. The adaptation to multiple drivers is possible because of the limited interference in the electrical activity between competing drivers. In a system with two stable drivers, tissue closer to the first driver exhibits electrical activity which can be closely approximated by the activity one would expect if the second driver was not present. This principle can be implemented in our algorithm as outlined in Fig. \ref{fig:multisearch} where the extension to tissues with multiple drivers is also described. 

\section{\label{sec:level1C} Results} 

Table I shows the results from the driver search algorithm. The prediction success is consistently high both for the single circuit and two circuit scenarios. Additionally, the prediction is reached after a small number of electrogram recordings, typically around 5 --6. For comparison, the probability of recording an electrogram on the re-entrant circuit at any given point in the tissue is just under 2\%.

\begin{table}[h!]
\centering
\def\arraystretch{1.6}
\begin{tabular}{ c | c | c }
\hline
Target Driver & Prediction Probability & Mean Jumps\\
\hline
\hline
One Circuit & 95.4\% & 5.0 $\pm$ 1.7 \\ \hline
Two Circuits, Circuit 1 & 94.8\% & 5.2 $\pm$ 2.3 \\
Two Circuits, Circuit 2 & 92.5\% & 6.1 $\pm$ 4.2
\end{tabular}
\label{tab:res}
\caption{The success rate of the algorithm described in Fig. \ref{fig:alg}. The results are generated from 1,500 simulations each of the single and double driver tissues. The prediction success corresponds to the classifier models outputting a final positive prediction which lies on a position which if ablated would terminate the re-entrant circuit. The mean number of jumps refers to the total number of electrogram grid recordings required before reaching a final prediction.}
\end{table}

The results indicate that the algorithm is capable at locating simple re-entrant circuits on randomly generated CMP tissues with excellent prediction rates from a small number of recording locations. These results support the possibility of using statistical learning techniques for locating AF drivers on heart tissue using indirect feature measurements, derived from the accessible electrograms.\\ 

\section{\label{sec:level1D} Discussion}

It is clear from the results in section \ref{sec:level1C} that a statistical analysis of electrograms can be used to extract information as to the location of re-entrant drivers in the CMP model, an idealised mathematical model of AF focusing on the discretised structure of the atrial myocardium and the electrophysiological action of fibrosis. 

In recent work, we have reviewed issues concerning the limited resolution of mapping technologies used during ablation, and have found that these limitations make it hard to identify distinct mechanisms of AF in clinical practice \cite{Roney2017}. If re-entrant circuits are to be ablated, terminating AF, these issues must be addressed. However, the methods proposed here may mitigate some of the mapping resolution issues that arise from single localised measurements by focusing on optimising a single prediction of the driver location based on a statistical analysis of multiple lower resolution measurements. Such an approach maximises accuracy by minimising errors due to local noise. 

\subsection{\label{sec:lims}Future Work}

In its current form, our driver mapping approach is not refined enough for use in clinical practice -- the simplifications in the CMP model do not accurately represent real electrical behaviour in the heart. 

Limitations in the current work come under three categories, (a) intrinsic limitations in the original CMP model, (b) limitations in our simplified implementation of the original CMP model, and (c) limitations arising in the data analysis and machine learning procedures. 

(a) The CMP model is not a fully realistic model of the atria. It represents a simplified 2D cylindrical topology of the myocardium and has all cells arranged in a consistent square lattice. The refractory period of cells are uniform and fixed, and the cell to cell conduction velocities at the microscopic level are constant, traversing one cell to cell coupling per time step. In this current form, the CMP model is not suited to studying some of the rate dependent effects typically studied in continuous models of AF. However, the CMP model's explicit focus on tissue anisotropy and the presence of fibrosis demonstrates that the key features of AF can spontaneously arise without the need for the extra detail studied in other computationally intensive models \cite{clayton2011}. Furthermore, inhomogeneities in refractory period can be implemented with small adjustments to the model. Future work could consider testing the proposed search mechanisms in this adjusted model

(b) In our implementation of the CMP model, we worked in the low noise limit at $\delta = 0$ where new re-entrant circuits could not form after the tissue had been initialised with one or two artificial circuits in the tissue. This was done for computational simplicity and is a small source of noise compared to the effects of stochastic coupling on wavefront dynamics. The exception is that non-zero $\delta$ does allow for small local fluctuations on short timescales as shown by part (d) of Fig. \ref{fig:rotorform}. In a simple re-entrant circuit searching algorithm, these fluctuations can occasionally be mistaken for circuits. However, these fluctuations dissipate after a single cycle of electrical activity. Therefore, a full implementation of the search algorithm could easily account for these fluctuations by recording electrical dynamics over two or more periods. This implementation of the CMP model only analysed tissues with simple, two fibre re-entrant circuits typically formed in the real CMP model from a single dysfunctional cell. In the full CMP model, re-entrant circuits can form consisting of multiple fibres and multiple dysfunctional cells. On a local level, this can lead to different electrical dynamics. However, the global electrical dynamics are very similar across the different types of re-entrant circuit. It is also rare for complicated re-entrant circuits to form, therefore, the simple circuit implementation represents the typical case that would be relevant in a realistic setting. 

(c) Finally, there are some limitations that arise during data analysis of the electrograms. Electrograms generated from the CMP model are typically cleaner than real atrial electrograms where consistent measurements across the tissue are not always possible. Despite this, the most important features such as the direction of electrical propagation across a set of electrodes can still be found accurately using clinical electrograms \cite{roney2014,cantwell2015}. By taking the gradient of feature values across the electrode probes in clinical practice, much of the same statistical information that has been used in this work can be calculated and analysed. 

Since electrograms were analysed using supervised machine learning methods, it is also clear that the amount of accurate clinical data required to train models will be severely restricted. In the short term this might appear to be a major limitation. However, other medical studies have recently shown the efficacy of using simulations to pre-train machine learning models for clinical use before refining the models using clinical data. This process of ``pre-training'' before the models are slowly improved through the acquisition of real data has had recent success applied to gene splicing by \citeauthor{Rosenberg2017} \cite{Rosenberg2017}. It is also important to consider whether any artifacts in the dynamics of the CMP model may have effected the success of our algorithm in a way that could not be applied in a more realistic setting. Of particular note is the perfect isotropy along muscle fibres which may lead to wavefronts traveling in the longitudinal direction being unrealistically uniform. Despite this concern, this particular artifact is more likely to hinder the success of machine learning algorithms than be a benefit since wavefront uniformity can limit the model's capacity to distinguish between different regions of the CMP tissue. On the whole, this should be seen as a secondary concern when adapting this work to more realistic settings.

As a final note, it is interesting to highlight suprising parallels between the CMP model and recent developments in the mechanistic understanding of AF. \citeauthor{hansen2015} \cite{hansen2015} have recently demonstrated that micro-anatomical re-entrant circuits forming in the transmural region with variable orientation, between the epicardium and the endocardium, can result in a complex variety of breakthrough patterns observed at the surface of the epicardium. The surface of the epicardium is typically what is mapped during ablation and for most electrophysiological studies of the atria.
These breakthrough patterns can appear as full rotors, partial rotors or concentric activity -- a direct explanation of observations that were previously seen as in direct conflict but which can now be explained from a single mechanism. \citeauthor{nattel2017} \cite{nattel2017} has suggested this may act as a unifying mechanism for the understanding of AF. Interestingly, the work by \citeauthor{hansen2015} specifically associates the emergence of AF with the formation of micro-anatomical re-entrant circuits at the edges of regions with high fibrosis, in direct agreement with the mechanism described by the CMP model. Although the 2D formulation of the CMP model is currently unable to investigate the formation of varying breakthrough patterns at the epicardial surface, a 3D implementation should be able to reproduce the key features of these observations.

\section{\label{sec:level1E} Conclusion}

Despite major research efforts focusing on the theoretical background and clinical understanding of atrial fibrillation in recent years, ablation success rates have remained disappointing since the 1990s. Major improvements will only be possible given significant developments in our mechanistic understanding of AF and new technological approaches to ablation. The methods demonstrated here apply the benefits of machine learning and couple it with the capacity for large scale simulations of cardiac electrical activity in cellular automata. Given the efficiency of the model, extending the 2D CMP tissue to a 3D structure should be computationally easy and investigated as a priority. The work presented here is a first step in a simplified theoretical model, that demonstrates a clear potential for locating re-entrant circuits with high success from electrograms alone.

\bibliography{ref}

\clearpage

\begin{appendices}
\section{\label{sec:ap1}Electrogram Simulation}
For the CMP model, electrograms can be simulated using the following equation: 

\begin{equation} \label{eqn: ECG}
V(X^{'}, Y^{'}) = \sum\limits_{X, Y} \frac{\Delta X(\nabla_{x}V) + \Delta Y(\nabla_{y}V)}{(\Delta X^2 + \Delta Y^2 + \Delta Z^2)^{\frac{3}{2}}},
\end{equation}

\noindent where $(X, Y)$ are the cell positions on the CMP tissue, $\nabla_{x}V = V(X,Y) - V(X-1,Y)$ and $\nabla_{y}V = V(X,Y)-V(X,Y-1)$ are the discretised gradients, $(X^{'}, Y^{'})$ is the position of the probe, $\Delta X = X - X^{'}$, $\Delta Y = Y - Y^{'}$, and $\Delta Z$ is the distance between the probe and tissue surface. $V(X,Y)$ is set to be the state of the cell, where a resting cell has a state value of 0 and an excited cell has a state value of 50 in arbitrary units. For a refractory cell, the state value is a linear interpolation between these two extremities. This action potential was chosen as it simplifies the process but it can be mapped back to a more realistic representation -- doing so has a negligible affect of analysis. 

\section{\label{sec:ap2} Electrogram Visualisation and Analysis}

The features generated from the electrogram data is listed in Table. II. For analysis, the gradient of these features is calculated across the multi-electrode grid described in Fig. \ref{fig:elecgrid}. The relative time at which the sampling of raw electrode data starts can be exploited to calculate an approximate direction of the wavefront across the electrode grid. 

\begin{table}
\label{tab:features}
\centering
\caption{The features used for electrogram analysis. For an in depth description of individual functions see the \texttt{scipy} and \texttt{numpy} documentation.\footnote{Documentation for both \texttt{scipy} and \texttt{numpy} can be found at \url{https://docs.scipy.org/doc/}} The features were primarilly chosen for ease of use, descriptive capacity and speed at which they could be generated for large quantities of data. Before being processed, the raw electrogram data was sampled to ensure consistent pre-processing. This was done by cropping the raw data between the first two maxima of the dominant fourier frequency. This ensured the output sample was a consistent length (one driver cycle period) and the sampling was started at the same point in the cycle. The time at which the crop is started relative to the initial recording time corresponds to feature 24 in the table. Note that since the initial recording time is arbitrary, this cannot be used as a feature for an individual electrode probe, but the gradient of the initial crop time across a multi--electrode probe can be measured. The sample data is represented by the label ``X'' in the table. The gradient of the sample data is denoted by ``g(X)''. Functions described by ``f\# - f\#*'' correspond to an operation involving other features listed in the table. For the Fourier features labelled with (*), the largest 9 frequencies and their corresponding amplitudes are calculated. }
\label{my-label}
\begin{tabular}{l|l|l|}

\cline{2-3}
                          & \textbf{Feature Name}                                                   & \textbf{Scipy Function}                                                                        \\ \hline
\multicolumn{1}{|l|}{1.}  & Maximum Value                                                           & numpy.max(X)                                                                                   \\ \hline
\multicolumn{1}{|l|}{2.}  & Minimum Value                                                           & numpy.min(X)                                                                                   \\ \hline
\multicolumn{1}{|l|}{3.}  & Amplitude                                                               & f1 - f2                                                                                        \\ \hline
\multicolumn{1}{|l|}{4.}  & Intensity                                                               & \begin{tabular}[c]{@{}l@{}}numpy.sum(\\             numpy.absolute(X))\end{tabular}            \\ \hline
\multicolumn{1}{|l|}{5.}  & Maximum Gradient                                                        & numpy.max(g(X))                                                                                \\ \hline
\multicolumn{1}{|l|}{6.}  & Minimum Gradient                                                        & numpy.min(g(X))                                                                                \\ \hline
\multicolumn{1}{|l|}{7.}  & Amplitude Gradient                                                      & f6 - f5                                                                                        \\ \hline
\multicolumn{1}{|l|}{8.}  & Max. Gradient Time                                                      & numpy.argmax(g(X))                                                                             \\ \hline
\multicolumn{1}{|l|}{9.}  & Min. Gradient Time                                                      & numpy.argmin(g(X))                                                                             \\ \hline
\multicolumn{1}{|l|}{10.} & \begin{tabular}[c]{@{}l@{}}Amplitude Gradient \\ Time\end{tabular}      & f9 - f8                                                                                        \\ \hline
\multicolumn{1}{|l|}{11.} & \begin{tabular}[c]{@{}l@{}}Number of 0V \\ Crossovers\end{tabular}      & \begin{tabular}[c]{@{}l@{}}numpy.argwhere(\\             g(X)[i] * g(X)[i+1] < 0)\end{tabular} \\ \hline
\multicolumn{1}{|l|}{12.} & \begin{tabular}[c]{@{}l@{}}First 0V Crossover \\ Time\end{tabular}      & numpy.min(f11)                                                                                 \\ \hline
\multicolumn{1}{|l|}{13.} & \begin{tabular}[c]{@{}l@{}}Largest Fourier \\ Frequencies*\end{tabular} & numpy.fft.rfftfreq(X)                                                                          \\ \hline
\multicolumn{1}{|l|}{14.} & \begin{tabular}[c]{@{}l@{}}Largest Fourier \\ Amplitudes*\end{tabular}  & \begin{tabular}[c]{@{}l@{}}numpy.absolute(\\             numpy.fft.rfft(X))\end{tabular}       \\ \hline
\multicolumn{1}{|l|}{15.} & Fourier Sum                                                             & numpy.sum(f14)[:10]                                                                            \\ \hline
\multicolumn{1}{|l|}{16.} & \begin{tabular}[c]{@{}l@{}}Relative Fourier \\ Amplitudes*\end{tabular} & f14[i] / f15                                                                                   \\ \hline
\multicolumn{1}{|l|}{17.} & Mean                                                                    & scipy.stats.describe(X)[2]                                                                     \\ \hline
\multicolumn{1}{|l|}{18.} & Skewness                                                                & scipy.stats.describe(X)[4]                                                                     \\ \hline
\multicolumn{1}{|l|}{19.} & Kurtosis                                                                & scipy.stats.describe(X)[5]                                                                     \\ \hline
\multicolumn{1}{|l|}{20.} & Maximum Time                                                            & numpy.argmax(X)                                                                                \\ \hline
\multicolumn{1}{|l|}{21.} & Minimum Time                                                            & numpy.argmin(X)                                                                                \\ \hline
\multicolumn{1}{|l|}{22.} & Amplitude Time                                                          & f20 - f21                                                                                      \\ \hline
\multicolumn{1}{|l|}{23.} & \begin{tabular}[c]{@{}l@{}}Variance Post \\ Minimum\end{tabular}        & numpy.std(X[f21:])                                                                             \\ \hline
\multicolumn{1}{|l|}{24.} & Sample Start Index                                                      & (See caption)                                                                                  \\ \hline
\end{tabular}
\end{table}

Feature visualisation is an exceedingly useful tool in enhancing the understanding of our system, informing the design of machine learning models and at various levels of code verification.

There are a vast array of standard feature visualisations available in machine learning libraries. However, these do not account for the spatial dependence of our electrograms. Therefore, we have principally relied on a custom visualisation which we have coined the ``vector feature map'' (VFM).

\begin{figure}
\centering
\includegraphics[width=\linewidth]{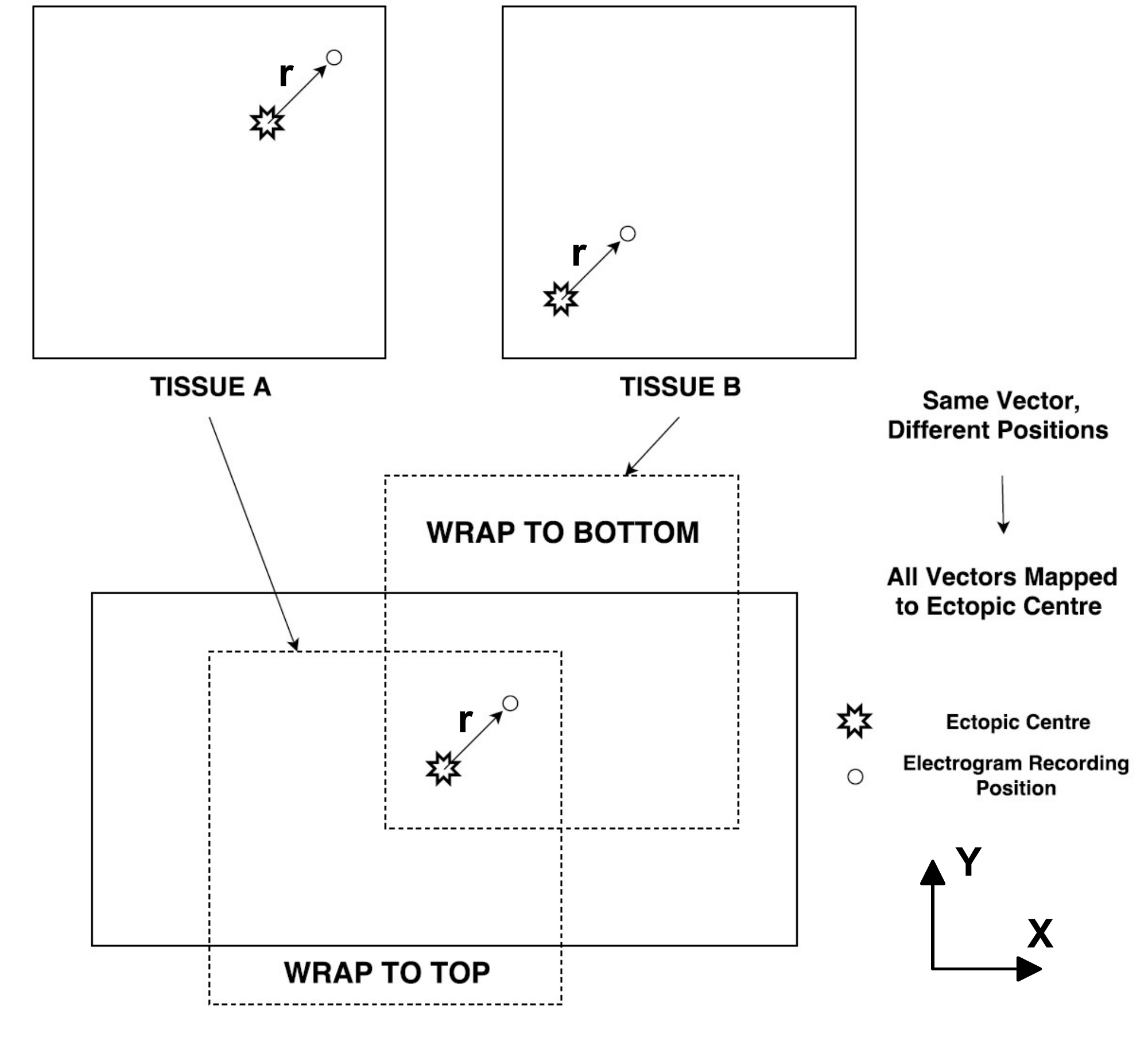}
\caption{Consider tissues $A$ \& $B$, where the drivers are located at coordinate ($150,150$) and ($50,50$), and electrograms are recorded at ($170,170$) and ($70,70$) respectively. Despite the electrograms being at different coordinates in the heart tissue, the relative vector from the driver to the electrogram is the same in both tissues, \textbf{r} $=$ ($20,20$). Hence, both electrograms typically record very similar wavefront behaviour. Note that not every tissue has electrograms recorded at every vector relative to the driver. The periodic boundary conditions of the CMP model in the $Y$ (transverse) direction mean that the relative vector from driver to recording position is always taken as the shortest distance between the two points -- this is expressed by the individual tissues being wrapped. This method used to generate vector feature maps. Thousands of instances (tissues) of the CMP model are generated with a single re-entrant circuit randomly positioned in each tissue -- two such tissues, $A$ \& $B$ are shown in this image. At a fixed vector away from the driver, \textbf{r}, we expect largely similar wavefront dynamics, and hence largely consistent electrogram features. These vectors can be mapped onto a single image by centering each tissue on an arbitrarily chosen fixed point of the tissue's re-entrant circuit. Each coordinate in the vector feature map then corresponds to the average value of a feature recorded at that particular vector away from the re-entrant circuit in thousands of different heart instances.}
\label{fig:fcmapping}
\end{figure}

In the simplified CMP model with a single artificial re-entrant circuit, each simulated heart tissue has the AF driver positioned at a different random position. Therefore, an electrogram recorded at the tissue centre is likely to show different behaviour in different heart instances. However, at a fixed vector away from the AF driver we would expect the general behaviour of features to be largely consistent independent of where in the heart tissue the driver is located. By simulating thousands of different heart instances, we can map the behaviour of features to a single image where each coordinate corresponds to the feature behaviour at that given vector away from the driver. This process is visualised in Fig. \ref{fig:fcmapping}. The value shown in a VFM is the average over many instances. As such, these behaviours are not necessarily seen in individual tissue instances -- the visualisations only indicate general trends. 

\begin{figure}
\centering
\includegraphics[width=\linewidth]{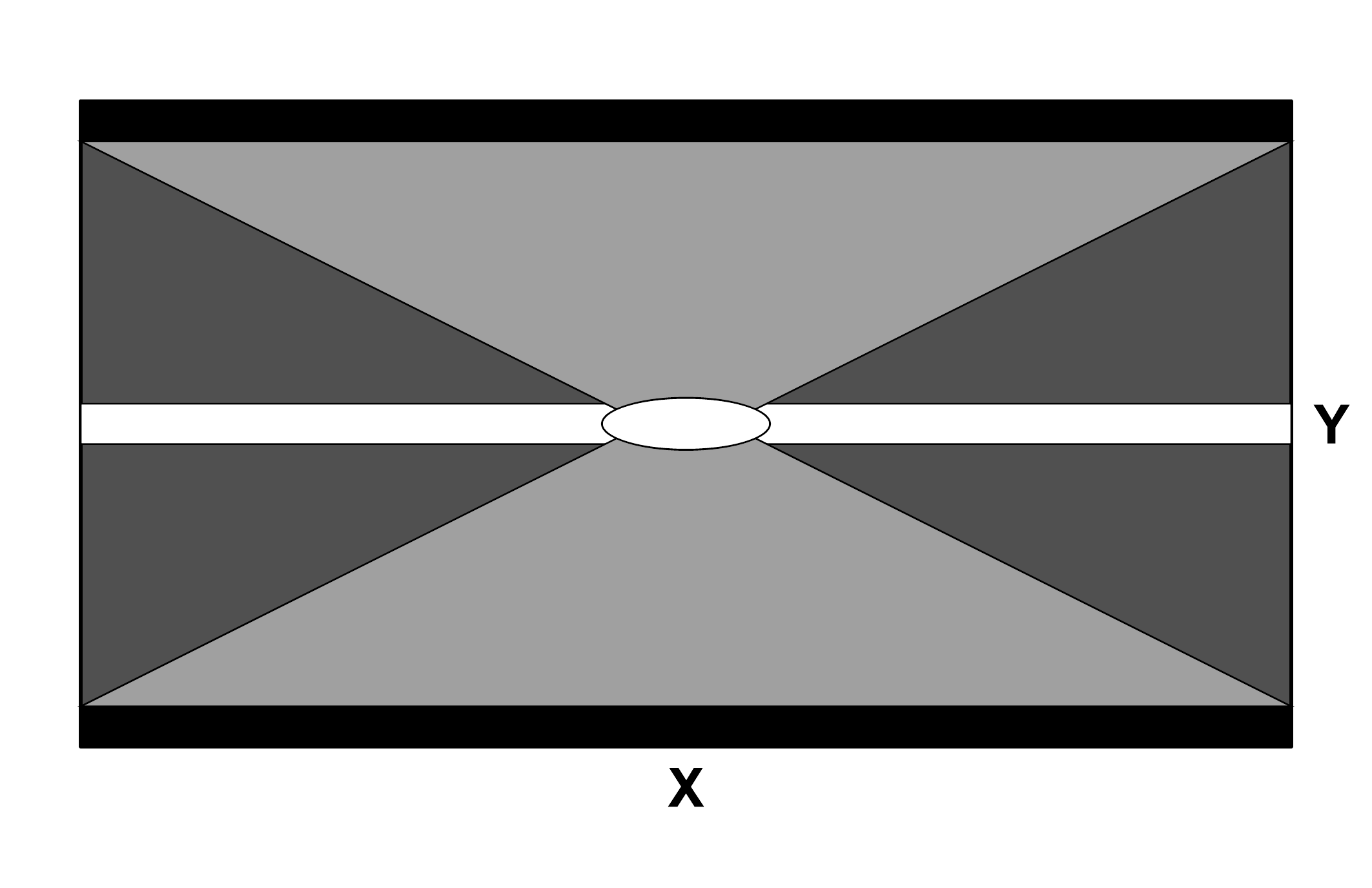}
\caption{A schematic showing the different regions on a vector feature map associated with different wavefront dynamics. The oval region at the centre corresponds to electrograms recorded near the AF driver -- wavefronts in this region spread in all directions. The white strips either side from the centre indicate convex wavefronts which propagate along the $X$ axis centred on the driver. The black strips correspond to positions near the circular boundary where wavefronts propagating in the $+Y$ and $-Y$ directions from the driver collide. Finally, the grey regions represent the ``bulk'', usually crossed by wavefronts that are approximately planar. The dark grey regions correspond to tissue which statistically only observes unidirectional flow in $X$, i.e, all electrograms in the dark grey region on the left only observe wavefronts propagating from the right. The light grey regions represent positions relative to the ectopic beat which can statistically observe bidirectional flow in $X$, i.e, positions in the light grey regions could see wavefronts approaching from either $+X$ or $-X$ over various tissue instances. The transition between these regions is not always sharp and varies over tissue instances. Note that not all vector feature maps are sensitive to all the different wavefront dynamics shown here.}
    \label{fig:fcdiag}
\end{figure}

The general regions on a VFM showing distinct flow dynamics are shown in Fig. \ref{fig:fcdiag}. The key conclusions from VFM analysis are:

\begin{itemize}
    \item There is poor feature separation in the bulk. Although there are some indicators differentiating between regions of unidirectional flow and bidirectional flow, this transition is not sharp across tissue instances. 
    \item There is strong feature separation between both the bulk and the ectopic beat's $X$ axis, and between the bulk and the axis along which wavefronts collide. This means these regions will be most susceptible to detection via statistical methods.
    \item Single electrode features are symmetric across both the $X$ and $Y$ axes of the driver. However, this symmetry can be broken by gradient features from the multi electrode grid. The sharp transition in $Y$ flow direction when crossing the $X$ axis is visible in these features and can be used to constrain the search region for driver detection.
    \item Even with gradient based features, regions in the bulk close to the driver are hard to distinguish from the bulk far from the driver. Hence, finding the driver from a single set of electrogram recordings will be difficult.
    \item The stochastic formation of vertical couplings in the CMP model means that the transition in $X$ flow direction when crossing the $Y$ axis of the driver is only sharp at the driver (as opposed to other position along the $Y$ axis which don't coincide with the driver).      
\end{itemize}

Our aim is not to find the driver in a single step but rather collate a small number of measurements to find the driver to a high degree of accuracy. Bearing that objective in mind, the considerations above suggest the following approach for designing a driver locating algorithm for the simplified CMP model. 

\section{\label{sec:ap3}Random Forests}

The Random Forest model is built using a set of decision trees. Each decision tree is created via the use of labeled example responses with associated electrogram features. For a general tree, the trees response space is split into $J$ non-overlapping regions distributed according to a certain metric. For a quantitative tree, the metric is the residual sum of squares given by
\begin{eqnarray}
\sum_{j=1}^{J}\sum_{i\in R_{j}}(Y_i - \hat{Y}_{R_j})^2,
\end{eqnarray}

\noindent where $Y_i$ are the individual response values over all response regions $R_j$, and $\hat{Y}_{R_j}$ is the average response in the region $R_{j}$ given by:
\begin{equation}
\hat{Y}_{R_j} = \frac{1}{\left |R_{j}  \right |}\sum_{R_{j}} Y_{R_j}.
\end{equation}

For a qualitative tree, the metric is instead the Gini index given by:
\begin{equation}
G = \sum_{k=1}^{K}\hat{p}_{jk}(1-\hat{p}_{jk}),
\end{equation}

\noindent where $K$ is the number of possible responses and $\hat{p}_{jk}$ is the fraction of responses $k$ in region $j$.

\begin{figure}
\centering
\includegraphics[width=\linewidth]{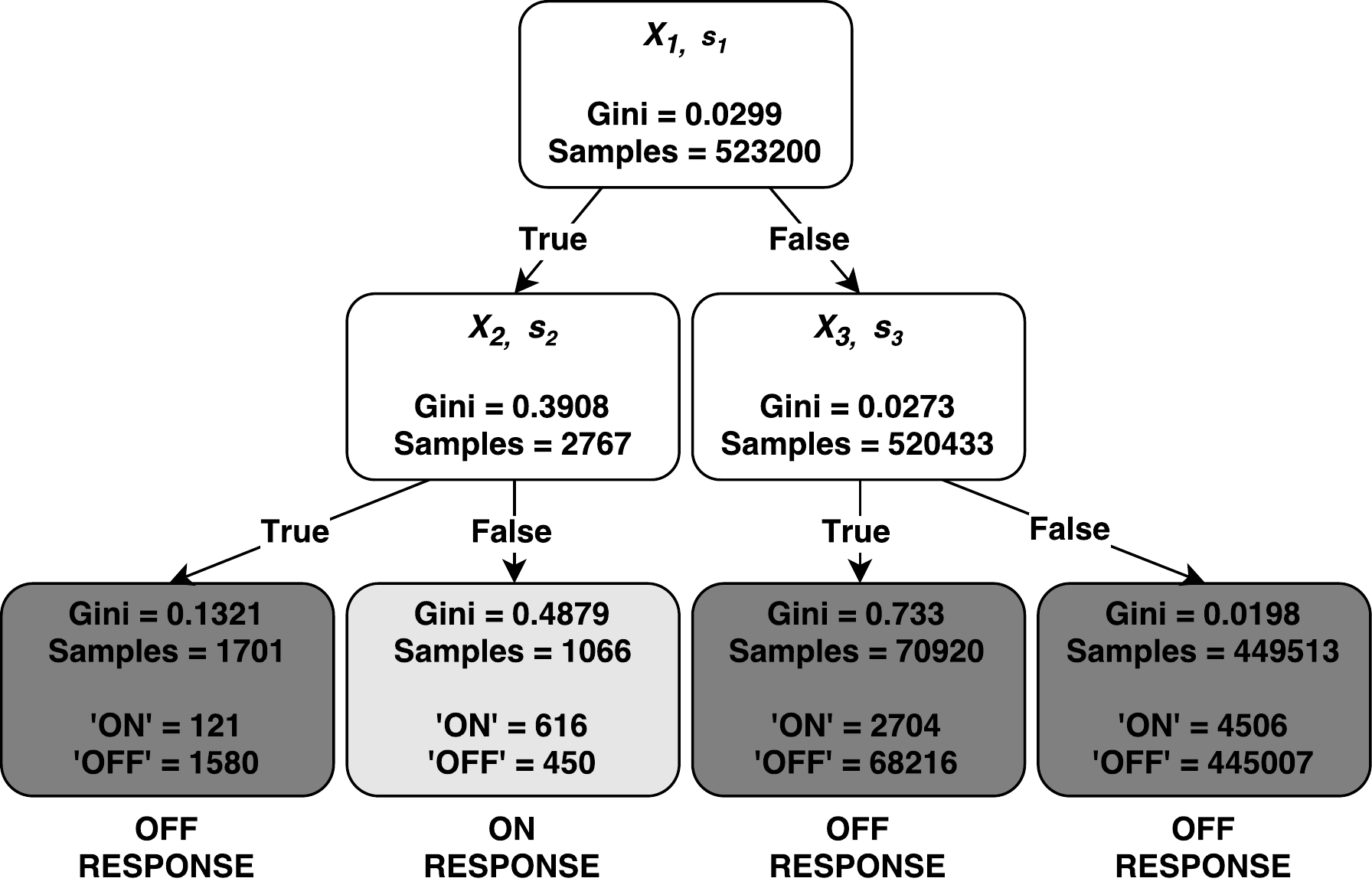}
\caption{Simple qualitative decision tree with a maximum depth of two. The responses are whether or not the probe is on or off the AF driver. Three features ($X_{1}, ~ X_{2}$ and $X_{3}$) with their corresponding thresholds ($s_1,~s_2$ and $s_3$) are used to create the response regions.}
\label{fig:decisiontree}
\end{figure}

The $J$ response regions are distributed by choosing an electrogram feature $X_n$ with threshold value $s$ which splits a region into the complementary regions $R_{1}(n,s) = \{X|X_n \geq s\}$ and $R_{2}(n,s) = \{X|X_n < s\}$. The feature and threshold are chosen to give the largest reduction in either of the chosen metrics at each split. This is a greedy top down process as the feature and threshold are chosen without considering future splits to simplify and speed up training, see Fig. \ref{fig:decisiontree}.

One of the drawbacks of decision trees is overfitting. When the tree is complex (large $J$), more features are used which reduces the bias but increases the variance of the response. This makes the decision tree overly specific to the training data, making the model very sensitive to small perturbations in the test data. Random Forests get around this through a process known as bagging. This process averages all the individual tree's responses thus reducing the variance but keeping the bias low -- this can be thought of as the mathematical equivalent of the expression ``the wisdom of the crowd''. For qualitative trees, a majority rule is used as the overall response. This effect is amplified by decorelating the individual decision trees by choosing the best feature from a random subset of the total features. The size of the random subset is $\sqrt[]{m}$ where $m$ is the total number of features processed from the electrogram waveforms.

The Random Forests built for this research were generated using the \texttt{scikit-learn} package in \texttt{python} with 15 decision trees \cite{scikit-learn}. This number of decision trees was sufficient to give good performance for locating the $X$ and $Y$ positions of the re-entrant circuit.

\section{\label{sec:ap4}Constraints \& Post-Processing of Predictions}

\begin{figure}[t!]
    \centering
    \includegraphics[width = \linewidth]{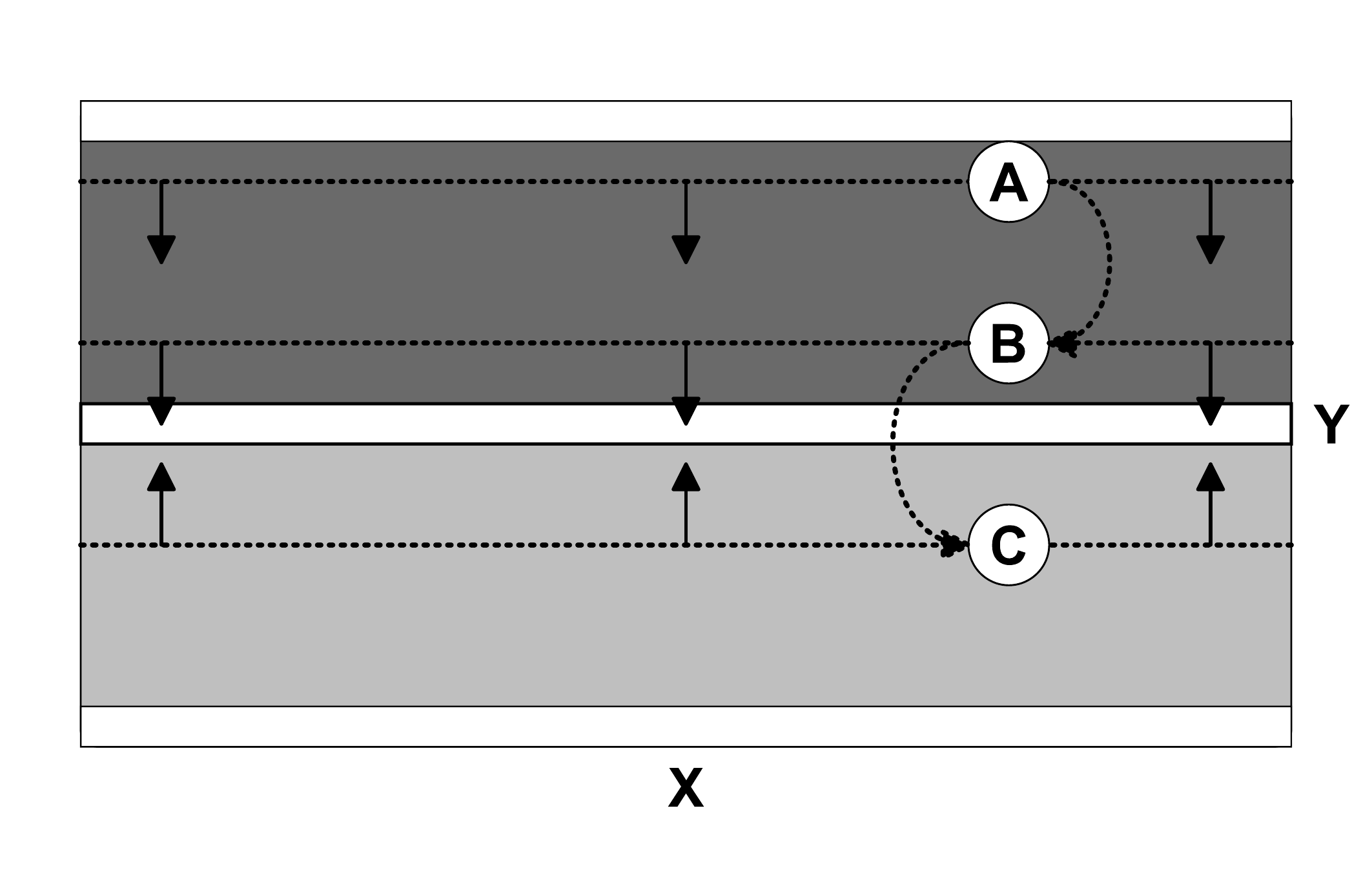}
    \caption{A schematic showing the method used to dynamically constrain the region searched for driver location by cross referencing flow information against the flow direction maps shown in Fig. \ref{fig:fcdiag}. This schematic uses the $Y$ flow map to constrain the $Y$ predictions. An equivalent (but less accurate) flow map can be generated for the $X$ direction. At each point in the CMP model, the net flow across the electrode grid can be calculated. This flow is strongly dependent on position relative to the recording position's closest driver. White regions indicate regions of zero net flow in the $Y$ direction corresponding to the driver axis (central white region) and the axis of wavefront collision (top and bottom white regions). The dark and light grey regions indicate net flow in the $+Y$ and $-Y$ directions respectively. For each set of electrogram recordings, the flow can be calculated. If $+Y$ flow is detected, the driver must be in the $-Y$ direction from the current position (cases $A$ \& $B$), and vice versa (case C). With each new electrogram recording the search area constrains towards the driver axis. }
    \label{fig:const}
\end{figure}

The individual Random Forest models trained can make independent predictions about the displacement of the current electrogram position from the driver. However, the full driver location algorithm must use these in synergy and restrict their predictions to known constraints. The algorithm is split into two phases, the $Y$ regression stage followed by the $X$ regression stage; it is shown in its full form in Fig. \ref{fig:alg}.

As a proof of concept, the constraints imposed here are kept reasonably simple. However, their importance should not be underestimated in ensuring the algorithms predictions converge to the true driver position and avoid infinite cycles of repeated prediction mistakes. 

\begin{figure}
\centering
\includegraphics[width = \linewidth]{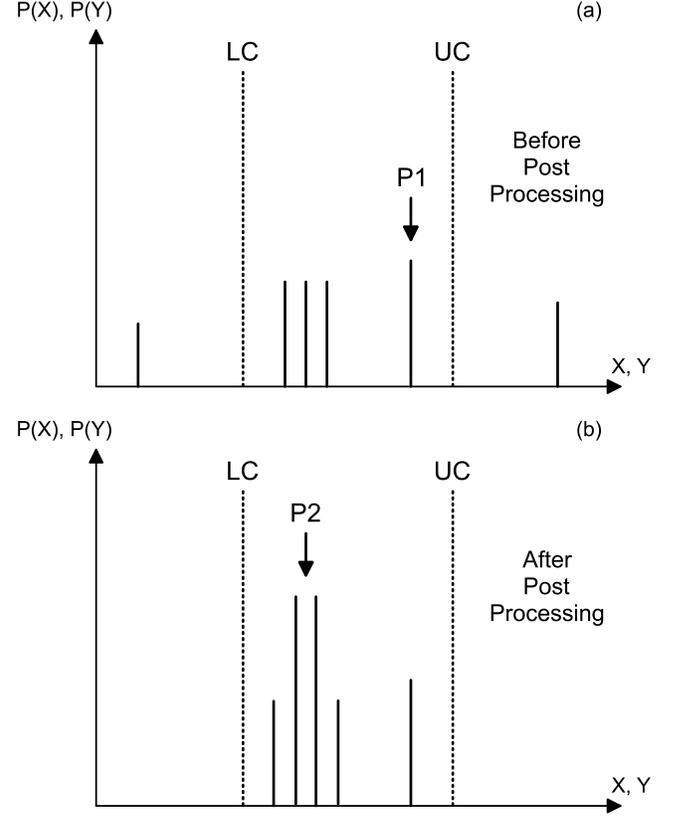}
\caption{A schematic showing the post processing of regression model predictions. Models output a probability distribution across the full range of $X$ or $Y$. The dotted lines represent upper and lower constraints (UC/LC). Probabilities outside the constraints are set to zero. A square wave convolution is then applied which sums the probabilities of predictions over a small range in $X$ or $Y$. This ensures large clusters of small probabilities are used for predictions (P2) instead of individual outliers with only marginally larger probabilities (P1). The square wave window is of width 2 and increases in size until a peak is found with $P > 0.5$. If the square wave reaches a width of 8 cells, the process is stopped and the output prediction is taken as the midpoint of the upper and lower constraints. Ignoring constraints, the convolution process improves regression predictions from an average error of 16.2 to 13.3 cells for $Y$ and from 6.4 to 4.7 cells in $X$.}
\label{fig:post}
\end{figure}

The first set of constraints use details in the change of local flow direction to dynamically restrict the search area. The mechanism is described in Fig. \ref{fig:const}. The predictions from individual regression models may not lie within the calculated constraints. Therefore, regression models are adapted to output a probability map for likely driver locations. The probabilities outside the constraint region are ignored and a square wave convolution is used to improve the predictions within the constraint window by giving a greater weight to a large cluster of slightly smaller probabilities rather than a single outlying larger probability. This process is illustrated in Fig. \ref{fig:post}.

In a multiple driver system, after the first driver has been found the search for the second driver is initiated as described in Fig. \ref{fig:multisearch}. The deviation of the wavefront collision point is used to constrain the search area to the areas marked in grey. For computational ease, our proof of concept insists that the algorithm cannot re-enter re-enter the region in which the previous driver was found. This is a small simplification and should not hinder the success of the algorithm in more complicated systems.  

\end{appendices}

\end{document}